\documentclass[12pt,preprint]{aastex}

\slugcomment{June 23 2003 ApJL in press}

\shorttitle{abundances diagnostics with \ion{Fe}{2}/\ion{Mg}{2}}
\shortauthors{Verner et al.}

\begin{document}


\title{Revisited abundance diagnostics in quasars: \ion{Fe}{2}/\ion{Mg}{2} ratios }


\author{E. Verner\altaffilmark{1,2,3} 
and F. Bruhweiler\altaffilmark{1,2,4}}
\altaffiltext{1}{IACS/Dept. of Physics, Catholic University of America.}
\altaffiltext{2}{Laboratory of Astronomy and Solar Physics, NASA/Goddard Space Flight Center Greenbelt MD 20771.}
\altaffiltext{3}{email: kverner@fe2.gsfc.nasa.gov}
\altaffiltext{4}{email: fredb@iacs.gsfc.nasa.gov}

\author{D. Verner\altaffilmark{1,5}}
\altaffiltext{5}{email: verner15@comcast.net}

\author{S. Johansson\altaffilmark{6}}
\altaffiltext{6}{Lund Observatory, Lund University, P.O. Box 43, S-22100 Lund, Sweden, sveneric.johansson@astro.lu.se}

\author{T. Gull \altaffilmark{2,7}}
\altaffiltext{7}{email: Theodore.R.Gull@nasa.gov}

\begin{abstract}
Both the  \ion{Fe}{2}~ UV emission in the 2000$-$3000~{\AA} region (\ion{Fe}{2}(UV)) and resonance emission line complex 
of \ion{Mg}{2} at 2800~{\AA} are prominent features in quasar spectra. The observed \ion{Fe}{2}(UV)/\ion{Mg}{2} emission ratios have 
been proposed as means to measure the buildup of the Fe abundance relative to that of the 
${\alpha}$-elements C, N, O, Ne and Mg as a function of redshift. The current observed ratios show large scatter 
and no obvious dependence on redshift. Thus, it remains unresolved whether a dependence on redshift exists 
and whether the observed \ion{Fe}{2}(UV)/\ion{Mg}{2} ratios represent a real nucleosynthesis diagnostic. 
We have used our new 830-level model atom for Fe$^+$ in photoionization calculations, reproducing 
the physical conditions in the broad line regions of quasars. This modeling reveals that interpretations 
of high values of \ion{Fe}{2}(UV)/\ion{Mg}{2} are sensitive not only to Fe and Mg abundance, but also 
to other factors such as 
microturbulence, density, and properties of the radiation field. We find that the 
\ion{Fe}{2}(UV)/\ion{Mg}{2} ratio combined with \ion{Fe}{2}~(UV)/\ion{Fe}{2}~(Optical) 
emission ratio, where \ion{Fe}{2}~(Optical) denotes 
\ion{Fe}{2} emission in 4000$-$6000~{\AA} band, can be used as a 
reliable nucleosynthesis diagnostic for the Fe/Mg abundance ratios for the physical 
conditions relevant to the broad-line regions of quasars.
This has extreme importance for quasar observations with the Hubble Space Telescope
 and also with the future James Webb Space Telescope.
\end{abstract}


\keywords{atomic processes---line: formation---methods: numerical---quasars: emission lines}


\section{INTRODUCTION}
The observed ratio of the ultraviolet  \ion{Fe}{2}~ UV emission flux 
(2000$-$3000~{\AA}; hereafter \ion{Fe}{2} (UV)) to that of the \ion{Mg}{2} resonance
doublet at 2800~{\AA} in broad-line regions (BLRs) of quasars has a strong potential for being 
a fundamental cosmological metallicity indicator.  In models of galactic
chemical evolution, the ratio of Fe to ${\alpha}$-element  (O, Ne, Mg) abundances
constrains the ages of star-forming systems (Wheeler et al. 1989). The Fe/${\alpha}$ age 
constraint follows from a different enrichment timescale, namely: ${\alpha}$-elements   are produced 
in supernova explosions of short-lived massive stars (primarily SN Type II), while
Fe has a large contribution from the longer-lived intermediate-mass binaries that 
produce Type Ia supernovae. Predictions indicate that the Fe/${\alpha}$ abundance ratio 
should increase by more than an order of magnitude for stellar systems older 
than 1 Gyr (Hamann \& Ferland 1993; Yoshii et al. 1996). However, if star 
formation has a significant contribution from very massive objects ($100 < M_{\star} < 1000 M_{\odot}$) 
as suggested recently (Heger \& Woosley 2002), then substantial Fe production 
can occur at times earlier than 1 Gyr. The answer to whether the Fe/${\alpha}$ abundance ratio
has a sharp break at 1 Gyr has important consequences not only for the evolution of
the elements, but possibly for understanding the formation of active galactic nuclei (AGN) as well. 
Whether the Fe/${\alpha}$ enrichment is due to SN Ia or SN II explosions is a key 
problem to be addressed in constructing our picture of galactic evolution in the Early Universe. 

The delay time in Fe enrichment can be estimated from observations
by plotting the observed \ion{Fe}{2} (UV)/\ion{Mg}{2} emission ratio (or the Fe/Mg abundance) as a function 
of redshift.  Recent \ion{Fe}{2} (UV)/\ion{Mg}{2} emission ratio measurements (e.g. Iwamuro et al.  2002; 
Dietrich et al. 2002; Freudling et al. 2003) show a wide range of values. That obscures any dependence on redshift. It is still unclear:
(1) whether a dependence on redshift exists; and 
(2) whether the derived observed \ion{Fe}{2}/\ion{Mg}{2} values are sensitive to other physical parameters besides Fe abundance. 

How the observed \ion{Fe}{2} (UV)/\ion{Mg}{2} emission ratio varies with physical conditions must be resolved
before one can develop a reliable model that accurately predicts the \ion{Fe}{2} emission line spectra in the
BLRs of luminous AGNs. The large velocity dispersion of \ion{Fe}{2} emission in BLRs, the large number of overlapping
lines, and the overall richness of the \ion{Fe}{2} spectrum (Verner et al. 1999)
form broad emission blends, producing a "\ion{Fe}{2} pseudo-continuum" 
and/or prominent features from the UV through visual (4000$-$6000~{\AA}; hereafter \ion{Fe}{2}~(Optical)) to IR. 
This necessitates a complete simulation 
of the physical processes affecting 
the Fe$^+$ ion to deduce how the \ion{Fe}{2} emission 
varies with properties of the radiation field, density, temperature, and Fe abundance in the emitting regions. 
To investigate what really makes \ion{Fe}{2} emission strong in the rest-UV, we have used a new Fe$^+$ model ion.
The model includes 830 atomic
levels for Fe$^+$ (up to $\approx 14~eV$) and represents an expansion of the previous model
of Verner et al. (1999). 

In this paper we evaluate how \ion{Fe}{2}(UV)/\ion{Mg}{2} 
and \ion{Fe}{2}(UV)/\ion{Fe}{2}(Optical) emission ratios vary due to extreme changes of the radiation field.
Then, for the given density and radiation field, we investigate the \ion{Fe}{2} emission properties 
within a reasonable range of Fe abundance and microturbulence. 

\section{EFFECTS OF PHYSICAL CONDITIONS ON \ion{Fe}{2} EMISSION}
\subsection{The expanded Fe$^+$ ion model}
Our earlier 371-level Fe$^+$ ion model (for energies below 11.6 eV), incorporated into the code 
CLOUDY\footnote{http://pa.uky.edu/$\sim$gary/cloudy}, 
predicts the emission fluxes of 68,635 \ion{Fe}{2} lines (Verner et al. 1999). This 371-level model, plus accurate knowledge 
of the radiation field in a low-density plasma successfully predicted \ion{Fe}{2} emission in the lower 
density environments of the Orion Nebula and Eta Carinae Ejecta (Verner et al. 2000; Verner et al. 2002). 
Comparison of observations with model results indicates that anomalously strong \ion{Fe}{2} emission lines can be 
the result from line pumping due to coincidence of \ion{Fe}{2} transitions with strong emission lines of other 
abundant ions (\ion{H}{1}, \ion{He}{2}, \ion{Si}{3}, \ion{C}{4}, etc.; Hartman \& Johansson 2000).

The higher radiation and particle density regimes in the BLRs of quasars favor the population of higher 
energy levels in Fe$^+$. Consequently, we expanded the original Fe$^+$ model to 830 levels 
(energies below 14.06 eV yeilding 344,035 transitions) and included more accurate atomic data -- new experimental 
energy levels (Johansson 2002, private communication)  and A-values from the Amsterdam 
database\footnote{ftp://ftp.wins.uva.nl/pub/orth/iron} (see also Raassen \& Uylings 1998). 
Where A-values were absent for many semi-forbidden transitions, and rate coefficients uncertain 
for {Ly$\alpha$} pumping channels and primary cascades, we have inserted reasonable estimates.

The complete simulation 
of the physical processes affecting the Fe$^+$ ion makes it possible to deduce the density, temperature, plus predict 
iron abundance in the emitting regions. To accurately model the \ion{Fe}{2} emission and Fe abundances, 
we must quantify the effects of the radiation field, and, only then, use \ion{Fe}{2}(UV)/\ion{Mg}{2} ratios to 
determine relative abundances. This approach is especially important for high redshift quasars 
where selection effects due to the luminosity may play a significant role. Thus, our knowledge 
about radiative excitation in Fe$^+$ is crucial to predict the \ion{Fe}{2} spectrum 
and Fe abundances at high redshifts. The questions is, "Can we use the observed \ion{Fe}{2}(UV)/\ion{Mg}{2} 
and \ion{Fe}{2}(UV)/\ion{Fe}{2}(Optical) ratios to derive Fe abundances?"




\subsection{Model parameters and comparison of the 830 and 371-level models} \label{bozomath_a}

In our modeling, we have chosen physical conditions where lines from low-ionization
species like \ion{Fe}{2} dominate the emission.  We adopt a range of Fe abundance
and a constant density, {$n_H$ }= 10 $^{9.5}$ cm$^{-3}$, with a total hydrogen column
density, $N_{H}$ = 10$^{24 }~$cm$^{-2}$. We further assume that the flux of hydrogen ionizing photons
at the illuminated face is 10$^{20.5}~$cm$^{-2}$ s$^{-1}$. These parameters for BLR conditions
are within the range of values taken from Verner et al. (1999), and Verner (2000). We employ the characteristic AGN
continuum described in Korista et al. (1997), which consists of a UV bump peaking
near 44 eV, a $f_{\nu} \propto \nu^{-1}$ X-ray power law, and a UV to optical spectral index, $a_{ox}=  -1.4$.
For the adopted parameters,
we have investigated how variations in abundance and microturbulence alter 
the intrinsic emission ratios of \ion{Fe}{2}(UV)/\ion{Mg}{2}  and \ion{Fe}{2} (UV)/\ion{Fe}{2}(Optical).

The differences in the predicted \ion{Fe}{2} spectrum 
calculated using 371 levels and 830 levels are negligible 
for lower densities and radiation field with depleted iron abundances (e.g. Orion Nebula). 
Differences in the predicted spectra from the two models become more significant with increased
radiation field and iron abundance. Moreover, the shape of the radiation 
field in AGNs leads to increased population of higher energy levels in \ion{Fe}{2}. 
Since densities are higher and the radiation field is more intense in BLRs, these upper 
levels provide additional excitation channels in the far UV (energies higher than 12 eV). 

Figure 1 compares the predictions using our 371- and 830-level models 
for the \ion{Fe}{2} pseudo-continuum appropriate for BLR physical conditions in QSOs. 
The plotted spectra are given as flux versus wavelength, where flux is ${\nu}f_{\nu}$, in units of erg~cm$^{-2}$~s$^{-1}$.  
The comparison clearly shows how the increased number of atomic levels 
produces a stronger \ion{Fe}{2} pseudo-continuum in QSO spectra. The 830-level ion model is far 
more accurate than the previous best efforts (Wills et al. 1985; Verner et al. 1999). The increased number of Fe$^+$ energy 
levels influences not only the \ion{Fe}{2} spectrum but lines of other elements and the whole energy budget also. 
However, the complete study of the new theoretical emission spectrum is not a subject of this paper.

Table 1 shows the predicted \ion{Fe}{2}(UV)/\ion{Mg}{2} and \ion{Fe}{2}(UV)/ \ion{Fe}{2}(Optical) ratios for
830-level and 371-level models of \ion{Fe}{2} versus 
contributions from the radiation field for a solar iron abundance and microturbulence of 1 km s$^{-1}$.
The models were calculated for radiative continuum pumping and for pure collisional excitation.
There is almost no change in predictions of
371-level and 830-level models for pure collisional excitation, if pumping by the radiation field is ignored.
However, the predicted ratios are strongly 
influenced by the radiation field and are sensitive to contributions from the upper levels in \ion{Fe}{2}.   
Figure 2 demonstrates the predicted \ion{Fe}{2} spectrum due to collisions and radiative pumping. 

\subsection{Effects of microturbulence and abundance} \label{bozomath_d}
How does the \ion{Fe}{2} emission depend upon microturbulence? 
The first successful models were developed by Netzer \& Wills (1983) who also demonstrated that
the \ion{Fe}{2} emission strength is sensitive to the turbulent velocitiy. 
This sensitivity was confirmed by modeling by Verner et al. (1999),
who showed that the predicted UV \ion{Fe}{2} emission indeed 
became stronger with increased microturbulence. To explore
how microturbulence affects the observed \ion{Fe}{2}(UV)/\ion{Mg}{2} measurements,
we calculated models with different microturbulence parameters ($v_{turb}~=~1,~5,~10,~100~$km~s$^{-1}$). 

Figure 3 shows \ion{Fe}{2}(UV)/\ion{Mg}{2} and \ion{Fe}{2}(UV)/\ion{Fe}{2}(Optical) ratios as a function of microturbulence. 
For small values of $v_{turb}$ and with solar abundance, the \ion{Fe}{2}(UV)/\ion{Mg}{2} ratio 
is smaller than that of \ion{Fe}{2}(UV)/\ion{Fe}{2}(Optical). The ratios are the same
at $v_{turb}~=~5$~km~s$^{-1}$. However,
 the \ion{Fe}{2}(UV)/\ion{Mg}{2} ratio grows faster than that of \ion{Fe}{2}(UV)/\ion{Fe}{2}(Optical)
as a function of microturbulence. 
The UV \ion{Fe}{2} transitions often arise from levels at high excitation energies, and
the density of levels  increases substantially above 10 eV. 
Thus, the UV transitions are closer in energy than those in the optical range,
and their strengths are more affected by microturbulence than those of optical lines. 
Although both the \ion{Fe}{2}(UV)/\ion{Mg}{2} and \ion{Fe}{2}(UV)/\ion{Fe}{2}(Optical) ratios increase with microturbulence,
dependencies are not identical. 
Our calculations demonstrate 
that the most reasonable values for \ion{Fe}{2}(UV)/\ion{Mg}{2} ratios correspond to $v_{turb}~=~5$~$-$~10~km~s$^{-1}$.

Is there any specific \ion{Fe}{2} feature, line or spectral 
index in AGN spectra that can be used to derive the iron 
abundance accurately? 
To probe the effect of varying abundance, we repeated the 
calculation with one and five times solar iron abundance, 
keeping the microturbulence constant at $v_{turb}~=~1~$km~s$^{-1}$.  
We define the iron abundance sensitivity ratio as 
(I(\ion{Fe}{2})$_5$ - I(\ion{Fe}{2})$_1$)/I(\ion{Fe}{2})$_1$
 and signify integrated fluxes I(\ion{Fe}{2})$_1$ and I(\ion{Fe}{2})$_5$ for abundances of solar 
and five times 
solar. Calculations with $v_{turb}~=~10~$km~s$^{-1}$ (Figure 4) 
show that this sensitivity ratio has the same slope as it has with $v_{turb}~=~1~$km~s$^{-1}$.

We find that the UV \ion{Fe}{2} lines are generally less sensitive 
to abundance than those in the optical range. Energy levels responsible for UV and optical lines
are not populated proportionally with increasing abundance. The structure of the Fe$^+$ ion favors the production
of more optical emission through cascades from upper levels.

Figure 5 shows how \ion{Fe}{2}(UV)/\ion{Mg}{2} and \ion{Fe}{2}(UV)/\ion{Fe}{2}(Optical) emission ratios 
change with iron abundance (1, 5, and 10 solar). 
The increase in \ion{Fe}{2}~(Optical) emission with the Fe abundance is stronger than that in \ion{Fe}{2} (UV)
emission. 
Comparisons of radiation and/or turbulence effects confirm that the abundance is not the only
factor that affects the \ion{Fe}{2} (UV)/\ion{Mg}{2} ratio. A factor of 5 in abundance increases the 
\ion{Fe}{2} (UV)/\ion{Mg}{2} ratio
by factor less than 2. By using \ion{Fe}{2} (UV)/\ion{Mg}{2} in combination with the
\ion{Fe}{2}(UV)/\ion{Fe}{2}(Optical) emission ratio, we can deduce abundances while accounting for
microturbulence and radiation field effects.  Table 2 shows predictions for models 
with different microturbulence ($v_{turb} = 1 - 100$~km~s$^{-1}$) and iron abundance (1~$-$~10 of solar). 
The columns are 
the following: microturbulence, abundance, and calculated \ion{Fe}{2}(UV)/\ion{Mg}{2} and 
\ion{Fe}{2}(UV)/ \ion{Fe}{2}(Optical) ratios.

\section{SUMMARY}
By applying the new 830-level model ion for Fe$^+$, which is incorporated into detailed photoionization 
calculations, we have probed the feasibility of using the strong \ion{Fe}{2} emission seen in quasar 
spectra to derive accurate Fe/Mg abundance enrichments. This \ion{Fe}{2} emission would be especially important in determining 
the enrichment as a function of redshift and the nature of star formation in the Early 
Universe. We conclude that abundance is not the only factor that makes \ion{Fe}{2} emission strong 
over a wide range of wavelengths and  is not even the strongest factor. Our modeling indicates 
that microturbulence also leads to preferential strengthening of the \ion{Fe}{2} emission in UV. 

On the basis of the calculations presented here, we make five basic conclusions:

1. In contrast to recent observational \ion{Fe}{2}/\ion{Mg}{2}  measurements, 
we emphasize the need for low-z quasar observations to calibrate detailed modeling as used here. 
Our model clearly demonstrates that a simple \ion{Fe}{2} template cannot 
be applied for all quasars, because the Fe II sensitivity 
to radiation and turbulence is much stronger than that of the \ion{Mg}{2} doublet. 
Only detailed modeling that includes the all these effects can 
provide more accurate Fe II templates.

2. Due to strong Fe II line overlap, we must adopt a wavelength band approach to get constraints on 
density, radiation field, and turbulence in BLRs. This approach 
is more complex, but the computational tools are readily available.

3. Further modeling, beyond that presented here, is necessary. Model comparisons with high S/N observations of quasars, 
and of even higher resolution observations of objects like $\eta$ Car, will produce better atomic 
parameters, and more detailed understanding of the effects of the radiation field in 
astronomical objects. Future work using improved atomic data to fully explore effects of the
radiation field is in progress.

4. These results clearly illustrate that the continuum of AGNs is 
greatly affected by Fe II. The spectrum of Fe II simply cannot be ignored. Understanding Fe II 
is crucial for obtaining realistic modeling of restframe AGN spectra from the UV through the 
IR. Studies in the expanded spectral range, much wider than used in any previous efforts, 
are required. 

5. We suggest that our model is to be applied to low-redshift quasars spanning both a wide luminosity range, 
and the 1,000~$-$~10,000~{\AA} restframe wavelength range. As the first step in this direction,
the STIS and NICMOS capabilities can be used immediately
to study low-redshift quasars.
The Near Infrared Spectrograph (NIRSpec) on board 
the James Webb Space Telescope (JWST) would make it possible to measure spectra from 0.6 
to 5 micron with resolving power in the range needed to obtain \ion{Fe}{2}(UV) and \ion{Fe}{2}(Optical) 
bands from low-redshift to high-redshift quasars.    

\acknowledgments

This research of EV has been supported, in part, through NSF grant 
(NSF - 0206150) to CUA, and HST GO-08098. EV acknowledge useful discussion with G. Ferland,
K. Kawara, and  Y. Yoshii. SJ is suported by a grant from the Swedish National Space 
Board. We wish to acknowledge the use of the 
computational facilities of the Laboratory for Astronomy and Solar 
Physics (LASP) at NASA/Goddard Space Center. We give special thanks 
to Keith Feggans, Don Lindler, and Terry Beck for their computer services support.
We thank unknown referee for suggestions that improved the paper.





\clearpage 

\figcaption [f1.ps]{\ion{Fe}{2} pseudo-continuum predicted by 371 (inset) and 830 level models for
Fe$^+$ in QSOs' BLR conditions (section 3.1). The 1000$-$7000~{\AA} wavelength range 
devided into small bins of 6~{\AA} (consistent with a ~ 500 km~s$^{-1}$ FWHM at 3600~{\AA}) 
and \ion{Fe}{2} flux calculated in each bin. 
The total 
\ion{Fe}{2} flux in range 2000$-$3000~{\AA} increased by factor 1.4 (Table 1) compared 
to the older model (Verner et al. 1999). The dashed line separates the \ion{Fe}{2} 
pseudo-continuum from strong lines in the 371-level model. The 830-level 
model predicts many strong \ion{Fe}{2} features in the 1600$-$3500~{\AA} range.} 

\figcaption [f2.ps]{\ion{Fe}{2} spectrum calculated in a radiative continuum pumping model (solid line)
and a pure collisional model (dotted line and inset). Broadening by a non-turbulent FWHM of 2,000~km~s$^{-1}$ 
(as observed for broad lines in the quasar 3C 273) has been used to enable comparison with
observations.}

\figcaption [f3.ps]{\ion{Fe}{2}(UV)/\ion{Mg}{2} and \ion{Fe}{2}(UV)/\ion{Fe}{2}(Optical) ratios as a function 
of microturbulence. See text for further discussion.}
 
\figcaption [f4.ps]{The general sensitivity of the \ion{Fe}{2} continuum  
to changes in abundance (for constant $v_{turb}~=~10~$km~s$^{-1}$). 
The \ion{Fe}{2}(UV) emission is less sensitive to abundance than \ion{Fe}{2}(Optical).}  

\figcaption [f5.ps]{\ion{Fe}{2}(UV)/\ion{Mg}{2} and \ion{Fe}{2}(UV)/\ion{Fe}{2}(Optical) ratios 
as a function of abundance.}

\newpage
\rotatebox{-90}{
\epsscale{0.8}
\plotone{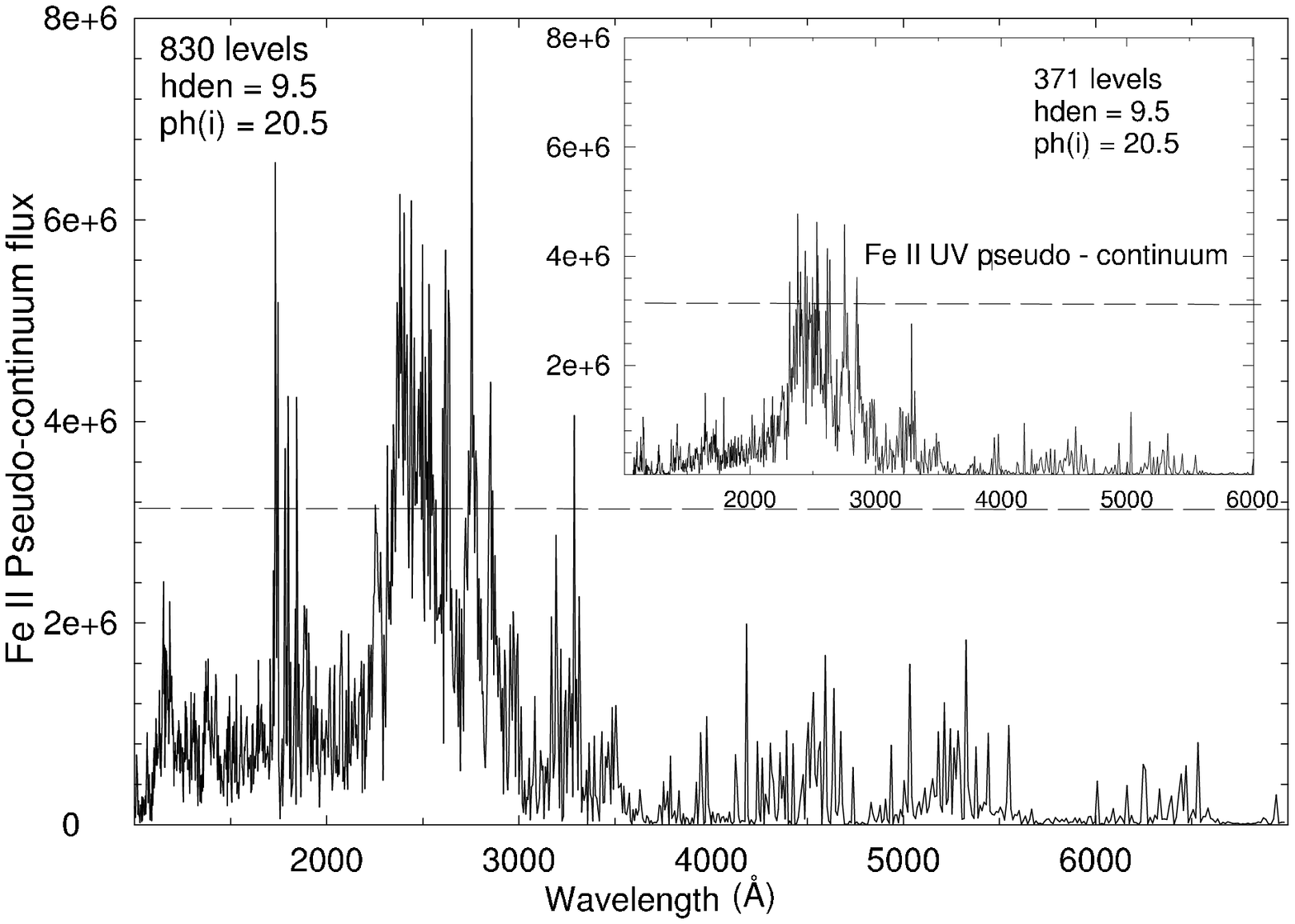}}
\rotatebox{-90}{
\epsscale{1}
\plotone{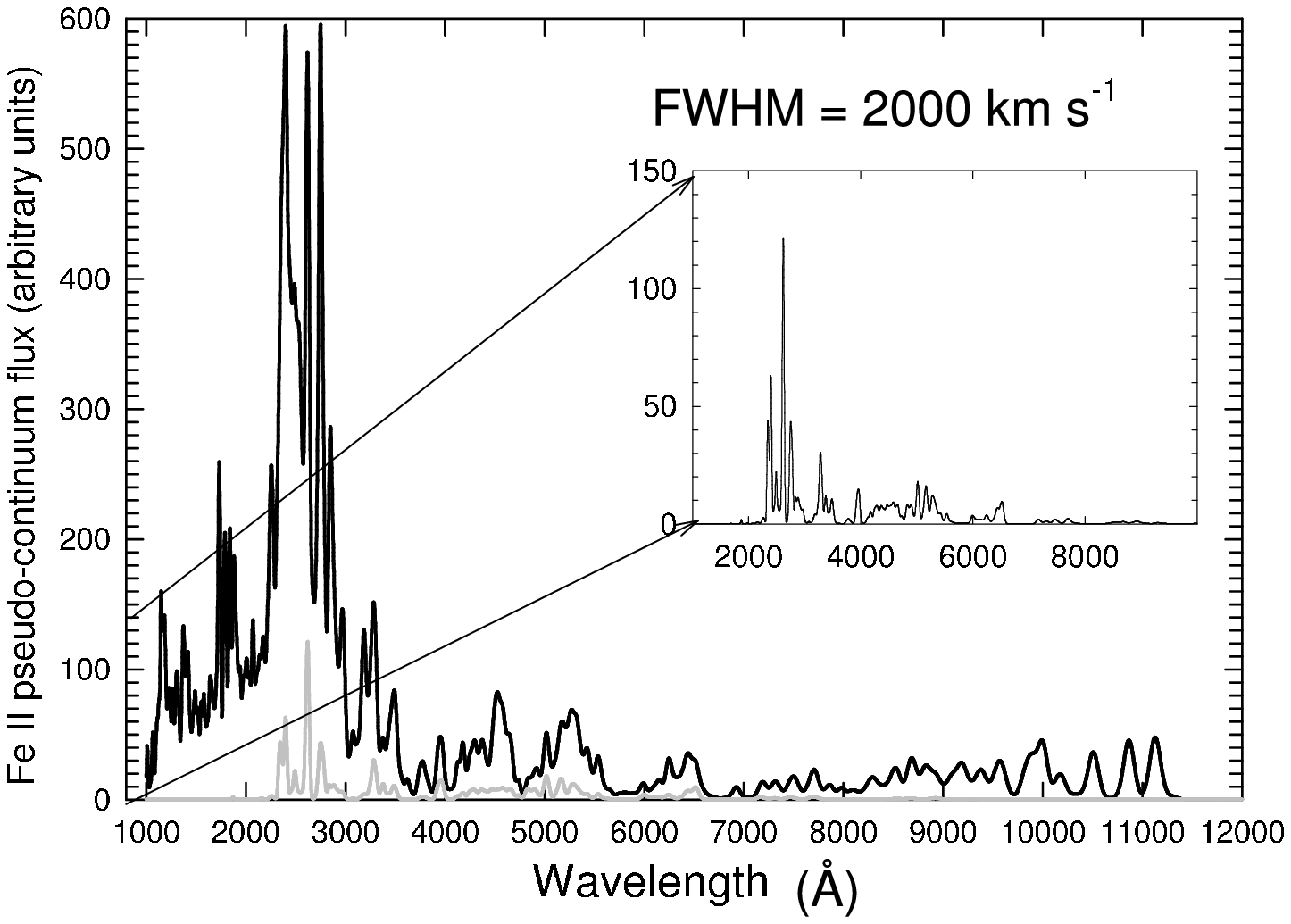}}
\rotatebox{-90}{
\plotone{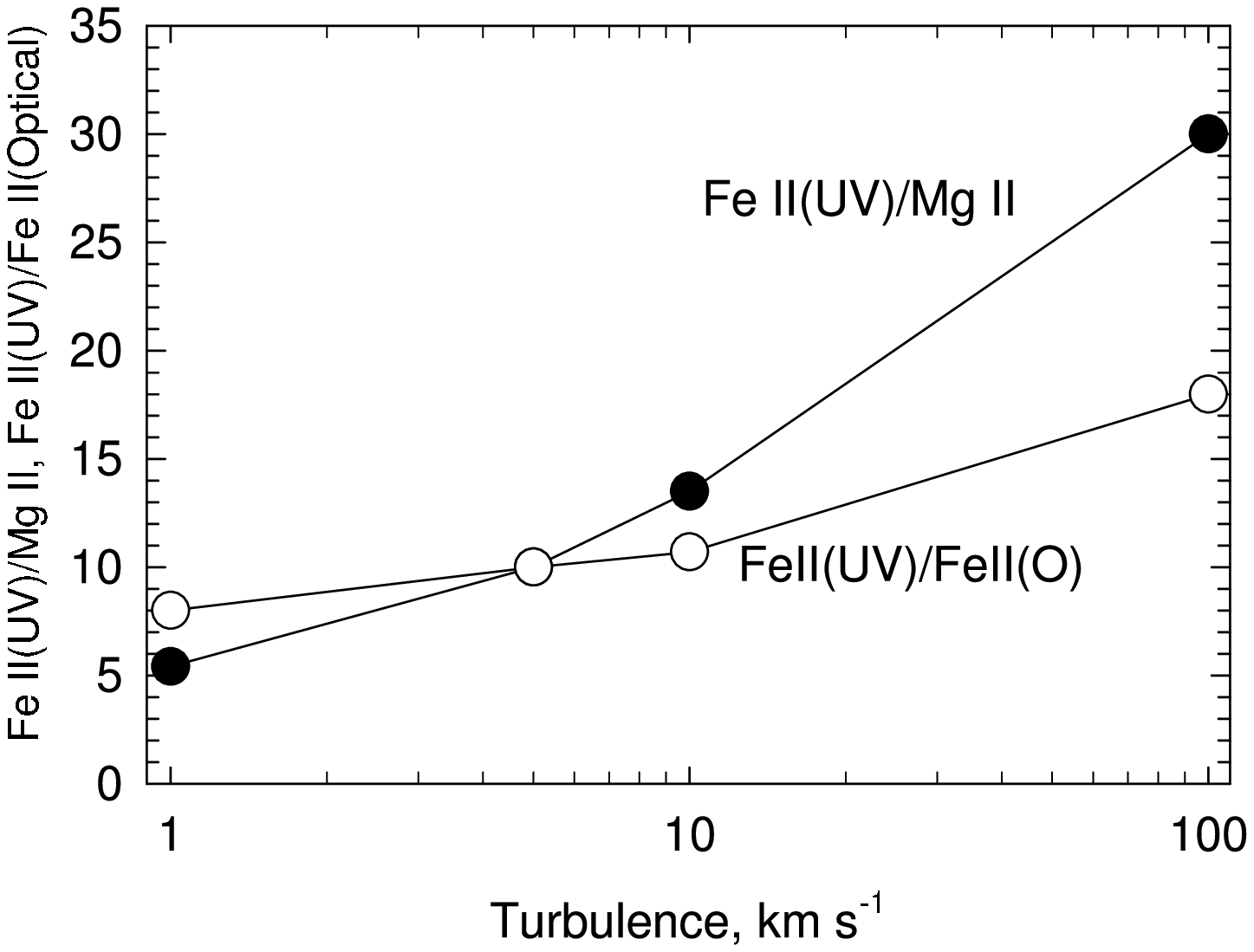}}
\rotatebox{-90}{
\plotone{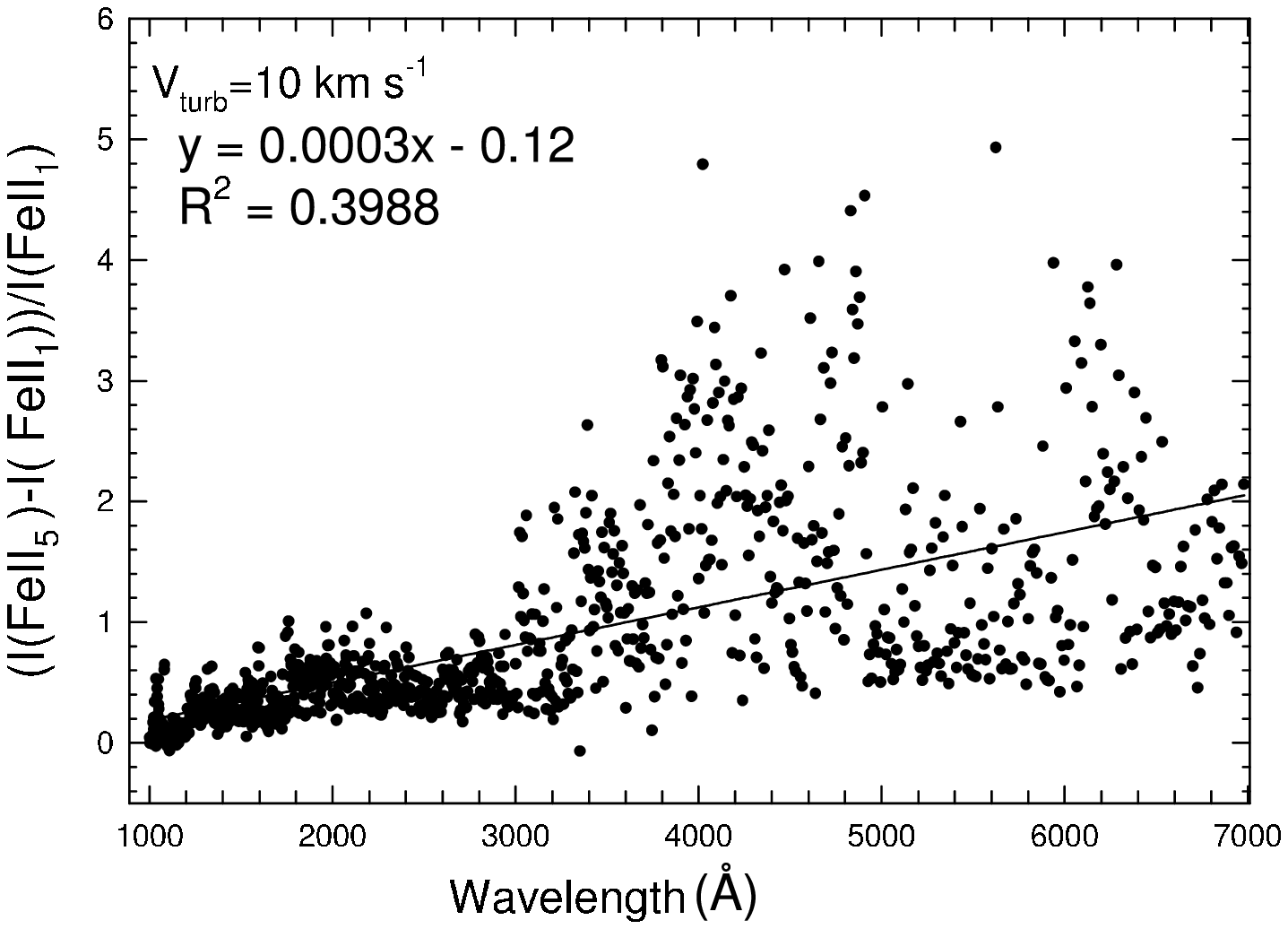}}
\rotatebox{-90}{
\plotone{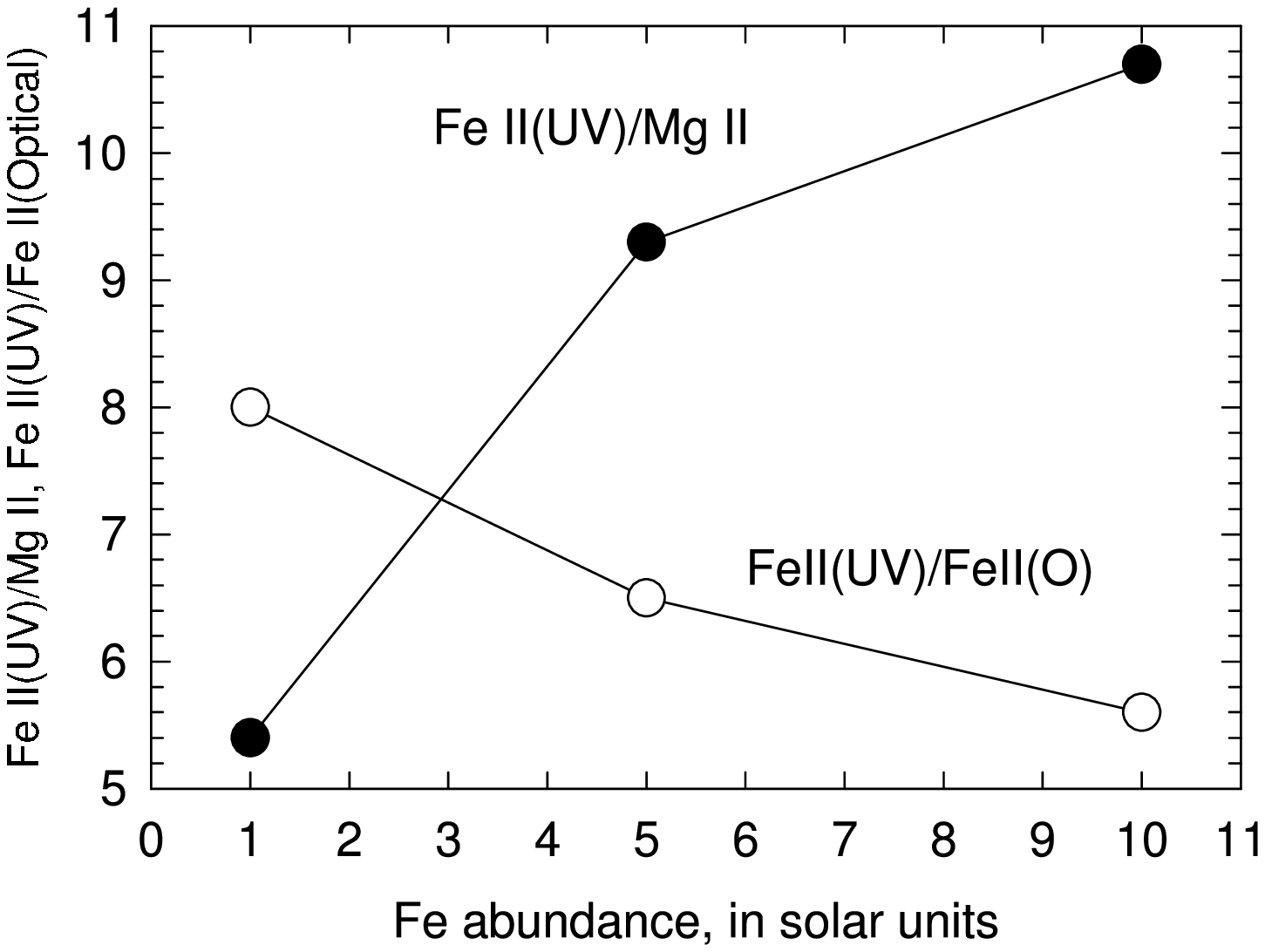}}






\clearpage

\begin{deluxetable}{cccc}
\tabletypesize{\small}
\tablecaption{Predicted \ion{Fe}{2}(UV)/\ion{Mg}{2} \& \ion{Fe}{2}(UV)/\ion{Fe}{2}~(Optical) Emission Ratios versus Excitation
for Solar Abundance and Turbulence of 1 km s$^{-1}$. \label{tbl-1}}
\tablewidth{0pt}
\tablehead{
\colhead{ Number of levels } & \colhead{Excitation}   & \colhead{Fe~II(UV)/Mg~II}   &
\colhead{Fe~II(UV)/Fe~II(Optical)} 
}
\startdata
371 & collisional & 0.5 & 2.8  \\
&  radiative & 4 & 11.4 \\
830 & collisional & 0.3 & 2.8  \\
&  radiative & 5.4 & 8 \\

\enddata




\end{deluxetable}

\clearpage

\begin{deluxetable}{cccc}
\tablecaption{Predicted \ion{Fe}{2}(UV)/\ion{Mg}{2} \& \ion{Fe}{2}(UV)/\ion{Fe}{2}~(Optical) Emission Ratios
versus Abundance and Turbulence. \label{tbl-2}}
\tablewidth{0pt}
\tablehead{
\colhead{ Turbulence (km s$^{-1}$)} & \colhead{Abundance}   & \colhead{Fe~II(UV)/Mg~II}   &
\colhead{Fe~II(UV)/Fe~II(Optical)} 
}
\startdata
& & & \\
1 & 1 & 5.4 & 8 \\
1 & 5 & 9.3 & 6.5 \\
1 & 10 & 10.7 & 5.6 \\
& & & \\
5 & 1 & 10 & 10 \\
10 & 1 & 13.5 & 10.7 \\
100 & 1 & 30 & 18 \\
& & & \\
5 & 5 & 14 & 6.8 \\
10 & 5 & 20 & 7.3\\
10 & 10 & 27.5 & 6.4 \\
100 & 10 & 77 & 9.2 \\
\enddata




\end{deluxetable}


\clearpage




\end{document}